\documentclass[aps,superscriptaddress,amsmath,amssymb,prl,twocolumn,10pt,floatfix]{revtex4}
\usepackage[dvips]{graphicx}
\usepackage{indentfirst}
\usepackage{bm}
\usepackage{enumerate}
\def\be{\begin{equation}}
\def\ee{\end{equation}}
\def\bea{\begin{eqnarray}}
\def\eea{\end{eqnarray}}

\begin{document}
\title{Dynamical instability of the XY spiral state of ferromagnetic
condensates}
\author{R. W. Cherng}
\author{V. Gritsev}
\affiliation{Physics Department, Harvard University, Cambridge, MA
02138}
\author {D. M. Stamper-Kurn}
\affiliation{Department of Physics, University of California,
Berkeley, California 94720}
\affiliation{Materials Sciences Division, Lawrence Berkeley National Laboratory, Berkeley,
CA  94720}
\author{E. Demler}
\affiliation{Physics Department, Harvard University, Cambridge, MA
02138}
%\author {$^2$ D. M. Stamper-Kurn}
%\affiliation{Department of Physics, University of California,
%Berkeley, California 94720}
\date{\today}
\begin{abstract}
We calculate the spectrum of collective excitations of the XY spiral
state prepared adiabatically or suddenly from a uniform
ferromagnetic $F=1$ condensate.
For spiral wavevectors past a critical value,
spin wave excitation energies become imaginary
indicating a dynamical instability.  We construct phase
diagrams as functions of spiral wavevector and quadratic Zeeman energy.
\end{abstract}
\maketitle

\begin{figure}
\includegraphics[width=3in]{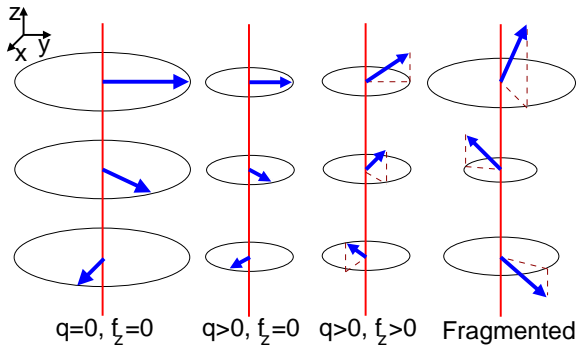}
\caption{
(color online) From left to right: magnetization vector in
the XY spiral state for fully polarized, partially polarized,
$f_z \ne 0$, and after fragmentation.
}
\label{fig:spiral}
\end{figure}

Spinor condensates of ultracold atoms
\cite{sengstock-05,chapman-05,bloch-06,dsk-06,bloch-07,pfau-05,ho-98,machida-98,yip-00}
are the latest addition to the class of many-body systems with
multicomponent order parameters.
%, that already includes liquid
%crystals, superfluid $^3$He, and magnets.
One of the most intriguing manifestations of the high symmetry of
such systems
%with spinful order parameters
%such systems
is the possibility of a large variety of spin textures and
topological defects. While earlier studies of liquid crystal
nematics and superfluid $^3$He demonstrated the existence of spin
textures, experiments with spinor condensates provide a unique
opportunity to investigate their non-equilibrium quantum dynamics.
%which should exhibit an intriguing interplay of charge and spin
%degrees of freedom.
Understanding dynamical properties of spin textures  will provide
valuable insight into many open problems of quantum magnets and
spinful superfluids, from analysis of the Kibble-Zurek mechanism of
nucleating topological defects when crossing a quantum phase
transition \cite{zurek-99} to finding the fundamental limits of
spinor BEC magnetometers \cite{higbie-07}.

In this paper we investigate theoretically the stability of the spin
spirals in ferromagnetic S=1 condensates (see Fig. 1.) Such states
represent the simplest type of spin structures and can be prepared
experimentally by applying a gradient of the magnetic field in the
direction perpendicular to the magnetization axis \cite{higbie-05}.
 Our main result is the
prediction of dynamical instabilities for spiral states, which we
summarize Fig. 2. We find that rotation of the magnetization vector
from the XY plane to the $z$ direction drives the instability at
small quadratic Zeeman energy while rotations within the XY plane
are responsible at large magnetic fields. Surprisingly we observe
that the unstable modes are characterized by wavevectors that can be
considerably larger than the wavevector of the initial spiral state.
Other instabilities in spinor condensates discussed previously
include Castaing instabilities \cite{castaing-84} in incoherent
non-condensed two component $^{87}$Rb \cite{kuklov-02} and the
modulational instability of a uniform spinor condensate
\cite{robins-01,ueda-05,lamacraft-07a}.
\begin{figure}
\includegraphics[width=3in]{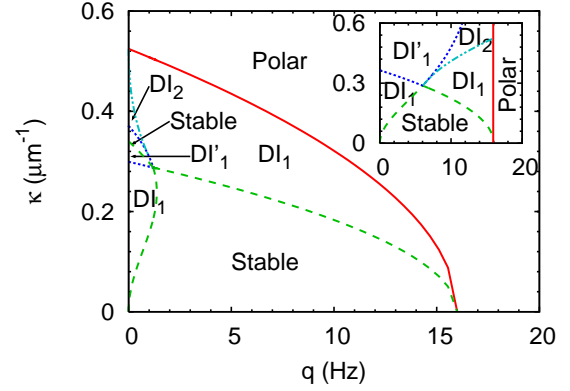}
\caption{
(color online) Collective mode phase diagrams for $f_z=0$ against
spiral wavevector $\kappa$ and quadratic Zeeman energy $q$ in the adiabatic
and sudden (inset) limits.
$DI_1$ $(DI'_1)$ indicates a dynamical
instability with one branch of unstable modes beginning at $k=0$ ($k>0$).  $DI_2$ indicates
a dynamical instability with two distinct branches of unstable modes.
}
\label{fig:critq}
\end{figure}

Our starting point is the microscopic Hamiltonian
\begin{equation}
\begin{split}
\mathcal{H}&=
\mathbf{\Psi}^\dagger\left[-\frac{\nabla^2}{2m}-\mu-pF_z+qF_z^2\right]\mathbf{\Psi}\\
&+\frac{g_0}{2} :\left(\mathbf{\Psi}^\dagger\mathbf{\Psi}\right) \left(\mathbf{\Psi}^\dagger\mathbf{\Psi}\right):
+\frac{g_s}{2} : \left(\mathbf{\Psi}^\dagger\mathbf{\Psi}^*\right)\left(\mathbf{\Psi}^T\mathbf{\Psi}\right):
\end{split}
\label{eq:hamiltonian}
\end{equation}
where $\mathbf{\Psi}_\alpha$ with $\alpha=x,y,z$ are annhilation operators for $F=1$ bosons with mass
$m$ and $(F_\alpha)_{\beta\gamma}$ are angular momentum operators.
We use a matrix notation with suppressed indices
where $*$, $T$, and $\dagger$ denote the complex conjugate,
transpose, and the conjugate transpose, respectively.
For example, $\mathbf{\Psi}$ ($\mathbf{\Psi}^\dagger$) is a column (row) vector
while $F_z$ is a matrix

Interaction strengths are given by $g_0=4\pi\hbar^2a_{0}/m$, $g_s=4\pi\hbar^2(a_{0}-a_{2})/3m$
\cite{ho-98}
in terms of the $s$-wave scattering lengths $a_F$ for two atoms colliding with
total angular momentum $F$ and $:\ :$ denotes normal ordering.
For $^{87}$Rb, $a_0=101.8 a_B$ and $a_2=100.4 a_B$ where $a_B$ is the Bohr radius \cite{verhaar-02}
giving positive $g_s$ and ferromagnetic interactions.

This Hamiltonian has a $U(1)\otimes SO(2)$ symmetry of global phase rotations and spin rotations
about the $z$ axis.
The chemical potential $\mu$ and linear Zeeman energy $p$ are Lagrange multipliers
controlling the corresponding conserved quantities
\begin{align}
\langle \mathbf{\Psi}^\dagger\mathbf{\Psi}\rangle&=n,&
\langle \mathbf{\Psi}^\dagger F_z\mathbf{\Psi}\rangle&=nf_z
\label{eq:conserved}
\end{align}
where $n$ is the total particle density and $f_z$ is the $z$ component
of the magnetization per particle.

Due to conservation of $F_z$, static magnetic fields enter through
the quadratic Zeeman energy $q$ instead of the linear Zeeman energy
$p$. Moreover, $q$ can be further manipulated through the AC Stark
shifts. From here on, we take representative values $q=70$ Hz
G$^{-2}$ $B^2$ where $B$ is the magnetic field and $n=2.2\times
10^{14}$ cm$^{-3}$ \cite{dsk-06}.  We neglect here magnetic dipole interactions
\cite{pu-04,saito-07}.

The XY spiral state is prepared from an initial cigar shaped
condensate with uniform XY magnetization by applying a magnetic field gradient
along the axial or $z$ axis.  After switching off the gradient, imaging of
the transverse magnetization can be used to study the stability of the XY spiral state.
As illustrated in Fig. \ref{fig:spiral},
the transverse magnetization winds along the $z$ axis.  The transverse
magnetization is fully polarized for $q=f_z=0$ and is suppressed
due to population of the $m_z=0$ component of $\mathbf{\Psi}$ when
$q\ne 0$ or $f_z\ne 0$.  In the presence of an instability, the
system fragments into domains carrying different magnetization vectors.

Analyzing the generation of the XY spiral state requires
understanding of the complicated non-equilibrium \textit{dynamics}.
We focus on studying the resulting non-equilibrium
\textit{stationary state} which we assume is well-described as a
coherent condensate. Such states are given by mean-field solutions
of the Gross-Pitaevskii (GP) equations implied by Eq.
\ref{eq:hamiltonian} which carry XY spiral order.  Compared to
stable ground states, these non-equilibrium stationary states are in
general metastable and decay via linear and non-linear processes. We
consider the linear stability of such states with respect to small
fluctuations by analyzing the spectrum of collective modes. The
distinction between stable and metastable stationary states also
arises for spinless bosons in a moving optical lattice
\cite{wu-01,smerzi-02} and in optics in the context of the four-wave
mixing instabilities (like the superradiance instability), in which
a mean-field treatment may yield a stationary situation only because
it neglects the spontaneous scattering into initially unoccupied
modes of the system \cite{fwm}.

To perform the stability analysis, it will be useful to consider a
frame comoving with the XY spiral order. We thus introduce the
substitution
\begin{equation}
\mathbf{\Psi}\rightarrow\exp\left(i\kappa z F_z\right)\mathbf{\Psi}
\label{eq:comoving}
\end{equation}
where $\kappa$ is the spiral wavevector.
The comoving frame
Hamiltonian is then given by Eq. \ref{eq:hamiltonian} with the subsitution
\begin{align}
p&\rightarrow p+\frac{i\kappa}{m}\nabla_z,&q&\rightarrow q+\frac{\kappa^2}{2m}.
\end{align}
We use this spin-dependent effective Hamiltonian to study the
stability of the non-equilibrium XY spiral state. However, we note
that it may be possible to explicitly engineer a physical
Hamiltonian of this form through continuous Raman excitation similar
to that described in Ref. \cite{higbie-02}.

In the adiabatic limit which we describe first, the components of
$\mathbf{\Psi}$ are able to adjust their populations in order
to accomodate the XY spiral order.
The interaction terms of Eq. \ref{eq:hamiltonian} describe
spin flip processes that mix the components of $\mathbf{\Psi}$
but still conserve the overall magnetization.  The components
of $\mathbf{\Psi}$ in the resulting XY spiral
state then describes a compromise between the kinetic energy cost of
the winding spiral and gain in interaction energy.

We begin by looking for mean-field solutions of the GP equations
in the comoving frame of the form $\mathbf{\Psi}=\sqrt{n}\mathbf{\Phi}e^{i\omega t}$
where we use the following parametrization
\begin{equation}
\mathbf{\Phi}=
\begin{bmatrix}
ie^{i\eta+i\eta_\perp}\cos(\phi+i\chi)\sqrt{\frac{f_z}{\sinh(2\chi)}}\\
ie^{i\eta+i\eta_\perp}\sin(\phi+i\chi)\sqrt{\frac{f_z}{\sinh(2\chi)}}\\
e^{i\eta}\sqrt{1-f_z\coth(2\chi)}
\end{bmatrix}
\label{eq:mf}
\end{equation}
which automatically gives the correct conserved quantities of
Eq. \ref{eq:conserved} by construction.

The parameter $\eta$ describes a global phase that
spontaneously breaks $U(1)$ phase rotation symmetry of the Hamiltonian.
Similarly, $\phi$ gives the orientation of the magnetization
vector and breaks $SO(2)$ spin rotation symmetry.
Here $\eta_\perp$ gives a relative phase between the $z$ and
transverse components of $\mathbf{\Psi}$.
Solutions of the GP equations for $\eta_\perp$
distinguish between $g_s>0$ ferromagnetic and $g_s<0$
antiferromagnetic interactions where $\eta_\perp=0$ and $\eta_\perp=\pi/2$,
respectively.  Recall we focus on the $g_s>0$ case.
Finally, $\chi$ controls the relative magnitude between
the $z$ and transverse components of $\mathbf{\Psi}$.
After introducing the dimensionless quantities
$\tau=\tanh(\chi)$ and $Q=q/2g_sn$ we find the GP equations give
the condition
\begin{equation}
Q\tau^3+\left(1-Q\right)\tau=f_z.
\end{equation}
As in Ref. \cite{ueda-07}, we find three classes of mean-field solutions:
polar, ferromagnet, and XY spiral state.
The polar state occurs for $f_z=0$ and $Q>1$ while the
ferromagnet occurs for $f_z=\pm 1$.  Both of these states occur only
on isolated lines in the mean-field phase diagram and do not
support XY spiral order.

We now analyze the stability of the XY spiral state
by studying the spectrum of collective fluctuations $\delta\mathbf{\Phi}$
about the mean-field solution $\mathbf{\Phi}$.
We take $\mathbf{\Psi}=\sqrt{n}(\mathbf{\Phi}+\delta\mathbf{\Phi})e^{i\omega t}$.
and focus first on the $f_z=0$ case.
Using the standard Bogoliubov analysis, we find the excitation
energies $\omega_k$ satisfies the eigenvalue equation
\begin{equation}
\det
\begin{bmatrix}
M_k-\omega_k&N\\
-N^*&-M_{-k}^*-\omega_k
\end{bmatrix}
=0
\label{eq:omegak}
\end{equation}
where $M_k$ and $N$ are given by
\begin{equation}
\label{eq:omegakaux}
\begin{split}
M_{k}=&
\frac{k^2}{2m}-\mu-\left(p+\frac{\kappa k_z}{m}\right)F_z+
\left(q+\frac{\kappa^2}{2m}\right)F_z^2+\\
&g_0n\mathbf{\Phi}^\dagger\mathbf{\Phi}+g_0n\mathbf{\Phi}\mathbf{\Phi}^\dagger+2g_sn\mathbf{\Phi}^*\mathbf{\Phi}^T\\
N=&g_0n\mathbf{\Phi}\mathbf{\Phi}^T+g_sn\mathbf{\Phi}^T\mathbf{\Phi}
\end{split}
\end{equation}
Recall we use a matrix notation so that
$\mathbf{\Phi}^\dagger\mathbf{\Phi}$ ($\mathbf{\Phi}\mathbf{\Phi}^\dagger$)
is a scalar (matrix).  We consider the one-dimensional case relevant
for cigar shaped condensates where we can take $k_z=k$.

\begin{figure}
\includegraphics[width=3in]{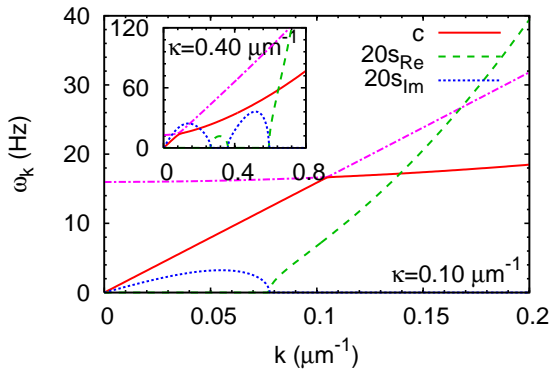}
\caption{
(color online) Representative collective mode dispersions for
$f_z=0$ and $q=0.2$ Hz in the adiabatic limit illustrating
one (two) branches of unstable modes. $c$ ($s$) denotes the
charge (spin) mode.
}
\label{fig:dispersion}
\end{figure}

The XY spiral state spontaneously breaks $U(1)\otimes SO(2)$ symmetry of global
phase and spin rotations and we find both a gapless charge and spin mode (as
required by the Goldstone theorem) with linear dispersions.
However, the spin mode can develop
imaginary frequencies which indicate the presence of a dynamical instability.

We find there are several distinct types of behavior for the spin mode
exhibiting a dynamical instability.
The first possibility is a branch of unstable modes starting at $k=0$
which we denote as $DI_1$ as illustrated in Fig. \ref{fig:dispersion}.
The second is a branch of unstable modes starting modes starting at $k>0$
which we denote as $DI_1'$.
The third possibility is two distinct branches of unstable modes starting
at $k=0$ and $k>0$ as illustrated in the inset of Fig. \ref{fig:dispersion}.

We construct the phase diagrams illustrated in Fig. \ref{fig:critq}
by characterizing the behavior of the spin mode as
a function of the spiral wavevector $\kappa$
and quadratic Zeeman energy $q$.
We first consider the adiabatic limit
characterized by an interpolation between long-wavelength
instabilities in the limit of large and small $q$.
Both instabilities can be throught of as unwinding of the spiral order
through rotations of the magnetization vector, but they
arise from qualitatively distinct physics.

When $q$ is zero, the system is rotationally symmetric.
In this limit, the XY spiral state is potentially unstable
towards unwinding through arbitrary $SO(3)$ rotations of the magnetization
vector from the XY plane to the $z$ axis.
However, small but finite $q$ provides a potential energy barrier
to such a process.  When the kinetic energy
stored in the non-uniform winding is sufficiently large,
small fluctuations corresponding to such rotations
can overcome this energetic barrier and grow exponentially
giving rise to a dynamical instability.  In particular,
the instability in this regime is due to fluctuations
in the \textit{direction} of the magnetization vector.

In contrast, large $q$ explictly breaks rotational symmetry.
The magnetization vector is essentially confined to the XY plane
and the unwinding of the XY spiral state can then only proceed
via $SO(2)$ rotations within that plane.  Such rotations proliferate
near the quantum phase transition to the polar state when fluctuations
in the \textit{magnitude} of the magnetization vector are large.

This large $q$ instability can be mapped to the
instability of current carrying states for spinless bosons \cite{altman-05}.
Here the effective $SO(2)$ magnetization order parameter maps
to the $U(1)$ order parameter of spinless bosons.
In addition, the critical fluctuations near
the transition to the polar state driving the instability
map to those of bosons near the Mott transition.

From the physical arguments above, we expect the XY spiral
state to be stable for wavevectors less than $\kappa^2/2m\sim q$
for small $q$ when kinetic energy is insufficient to overcome the
potential energy barrier.
Similarly, the XY spiral should be stable for wavevectors less
than $\kappa^2/2m\sim (q-q_c)$ near the quantum phase transition to
the polar state at $q_c$.
The boundaries in Fig. \ref{fig:critq} can be obtained explicitly
\cite{cherng-07} as
\begin{align}
\frac{\kappa^2}{2m}&\le\frac{2g_sn-q}{3+2\frac{g_s}{g_0}},&
q&\ge\frac{\kappa^2}{2m}\left(\frac{g_sn-\frac{\kappa^2}{2m}}{g_sn+\frac{\kappa^2}{2m}}\right)
\label{eq:cboundary_ad}
\end{align}
which gives $\kappa^2/2m\le q$ and $\kappa^2/2m=(q_c-q)/3$ for
for small and large $q$, respectively, in agreement with the
physical arguments.

Also notice in Fig. \ref{fig:critq} an isolated line on which the XY spiral
state is stable at small $B$ and intermediate $\kappa$.  In fact,
the dynamical instability in the $DI_1'$ surrounding this line is weak in
the sense that the imaginary part of $\omega_k$ is comparatively small.
The energetic arguments given earlier for the small $B$ limit seem to suggest
increasing $\kappa$ makes the XY spiral state \textit{more} unstable.
However, as $\kappa$ increases further, the populations in the components
of $\mathbf{\Psi}$ change appreciably.

In this limit, the XY spiral state can thought of as having a significant
polar component.  It has been shown previously that the polar state
has a dynamical instability with a characteristic wavevector
$k\sim\sqrt{2mg_sn}$ \cite{robins-01}.  When the spiral wavevector is on the order
of this characteristic wavevector, the XY spiral state
becomes \textit{less} unstable.  Physically, the spiral order
tends to suppress the instability of the polar component.

\begin{figure}
\includegraphics[width=3.4in]{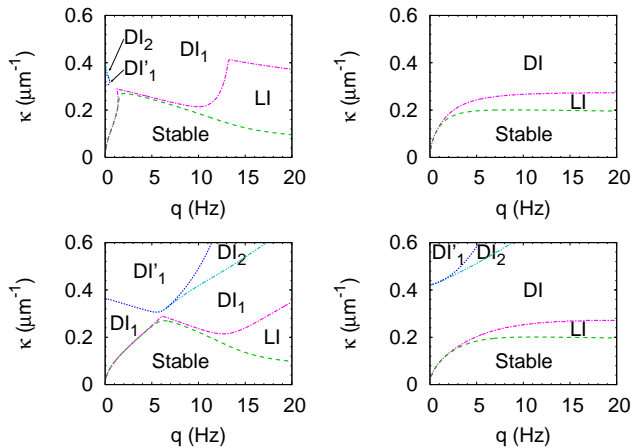}
\caption{
(color online) Collective mode phase diagrams for $f_z=0.05$ (left) and
$f_z=0.5$ (right) against
spiral wavevector $\kappa$ and quadratic Zeeman energy $q$ in the adiabatic (top)
and sudden (bottom) limits.
$DI$ ($LI$) indicates a dynamical (Landau) instability.
$DI_1$, $DI_1'$, and $DI_2$ are described in Fig. \ref{fig:critq}.
}
\label{fig:critq_multi}
\end{figure}

After analyzing the $f_z=0$ case in detail, we now briefly discuss
the $f_z\ne 0$ case.  The collective mode phase diagrams for the adiabatic
limit are shown in the top of Fig. \ref{fig:critq_multi} with
$f_z=0.05$ to the left and $f_z=0.5$ to the right.  The small $f_z$
phase diagrams are qualitatively similar to the $f_z=0$ case.
However, there is no polar state which only occurs for $f_z=0$
but there is an additional region which exhibits a spin mode
exhibiting a dispersion with negative frequencies.
corresponding to a Landau instability.  For large $f_z$,
the phase diagrams no longer exhibit a characteristic peak
for the stable region as a function of $q$.

So far the results have been focused on the adiabatic limit
where the components of $\mathbf{\Psi}$ can adjust due
to magnetization spin flip processes.  In practice, the
preparation of the XY spiral state can also occur on a
timescale shorter than that of spin flips.
We thus briefly comment on qualitatively similar results
in the sudden limit where the populations of each component
cannot change from their initial values.
To take this effect into account, we consider
mean-field solutions of the GP equation of the form
$\mathbf{\Psi}_\alpha=\sqrt{n}\mathbf{\Phi}e^{i\omega_\alpha t}$
with $\omega_x=\omega_y\ne\omega_z$ and $\mathbf{\Phi}$ is as in Eq. \ref{eq:mf}.
Notice the components of $\mathbf{\Psi}$ evolve
at different frequencies which allows for solutions
with the necessary populations for each component.

We then perform the same analysis of the collective modes as
in the adiabatic limit.  This gives for $f_z=0$ in the sudden limit
the phase diagrams in the inset of Fig. \ref{fig:critq}.
As in the adiabatic limit, the sudden limit is characterized
by an interpolation between long-wavelength instabilities
in the small and large $B$ limits arising from the
same physical origins.
The region where the XY spiral state is stable is given by
\begin{align}
\frac{\kappa^2}{2m}&\le\frac{2g_sn-q}{2+2\frac{g_s}{g_0}},&
q&\ge\frac{\kappa^2}{2m}\left(\frac{2g_sn}{g_sn+\frac{\kappa^2}{2m}}\right)
\label{eq:cboundary_sud}
\end{align}
which gives $\kappa^2/2m=q/2$ and $\kappa^2/2m=(q_c-q)/2$ for
the small $q$ and large $q$ limits, respectively.  Up to coefficients,
this is of the same form as the adiabatic limit.
The $f_z=0.05$ and $f_z=0.5$ phase diagrams for the
sudden limit are shown in the bottom of Fig. \ref{fig:critq_multi} and
exhibit the same structure as the adiabatic limit.

In this paper we have focused on the one-dimensional limit relevant
for cigar shaped condensates.  However, the formalism we used
can be readily adapted for the three-dimensional case.  In particular,
one just needs to take $k_z=k\cos\theta$ where $\theta$ is the angle between the
mode wavevector and spiral wavevector in Eq. \ref{eq:omegakaux}.

In summary, we studied a possible mechanism for the
instability of the XY spiral state. Focusing on the limits where the XY spiral is
prepared adiabatically or suddenly, we demonstrated that when the
spiral wavevector exceeds a critical value spin wave energies
become imaginary.  This indicates the presence of a dynamical instability
and exponential growth of fluctuations.  We traced the physical origin
of these instabilities to unwinding of the magnetization vector
through rotations from the XY plane to the $z$ axis for small quadratic Zeeman
energy $q$ and within the XY plane for large $q$.

This work was supported by NDSEG and NSF Graduate Research
Fellowship, Harvard-MIT CUA, AFOSR, MURI, and the NSF grant DMR-0132874.
When this work was being completed we learned of the work by A.
Lamacraft addressing related issues \cite{lamacraft-07b}.

\bibliography{spinor}
\end{document}